\documentclass{article}

\usepackage{arxiv}

\usepackage[utf8]{inputenc} 
\usepackage[T1]{fontenc}    
\usepackage{hyperref}       
\usepackage{url}            
\usepackage{booktabs}       
\usepackage{amsfonts}       
\usepackage{amssymb}  
\usepackage{nicefrac}       
\usepackage{xcolor}         
\usepackage{microtype}      
\usepackage{graphicx}
\usepackage{doi}
\usepackage{multirow}
\usepackage[toc,page]{appendix}

\usepackage{amsmath}
\DeclareMathOperator{\Rips}{Rips}
\DeclareMathOperator{\Alpha}{Alpha}
\DeclareMathOperator{\Conv}{Conv}
\DeclareMathOperator{\FC}{FC}
\DeclareMathOperator{\TF}{Transformer}

\DeclareMathOperator*{\Concat}{Concat}

\title{Multimodal Pre-Training Model for Sequence-based Prediction of Protein-Protein Interaction}

\author{ Yang Xue \\
	Baidu Inc. \\
        Shenzhen, China \\
	\texttt{xueyang02@baidu.com} \\
	\And
	Zijing Liu \\
	Baidu Inc. \\
        Shenzhen, China \\
	\texttt{liuzijing01@baidu.com} \\
        \And
        Xiaomin Fang \\
	Baidu Inc. \\
        Shenzhen, China \\
        \texttt{fangxiaomin01@baidu.com} \\
        \And
        Fan Wang \\
	Baidu Inc. \\
        Shenzhen, China \\
        \texttt{wangfan04@baidu.com} \\
}

\begin{document}
\maketitle

\begin{abstract}
Protein-protein interactions (PPIs) are essentials for many biological processes where two or more proteins physically bind together to achieve their functions.
Modeling PPIs is useful for many biomedical applications, such as vaccine design, antibody therapeutics, and peptide drug discovery.
Pre-training a protein model to learn effective representation is critical for PPIs.
Most pre-training models for PPIs are sequence-based, which naively adopt the language models used in natural language processing to amino acid sequences.
More advanced works utilize the structure-aware pre-training technique, taking advantage of the contact maps of known protein structures.
However, neither sequences nor contact maps can fully characterize structures and functions of the proteins, which are closely related to the PPI problem.
Inspired by this insight, we propose a multimodal protein pre-training model with three modalities: sequence, structure, and function (S2F).
Notably, instead of using contact maps to learn the amino acid-level rigid structures, we encode the structure feature with the topology complex of point clouds of heavy atoms. It allows our model to learn structural information about not only the backbones but also the side chains.
Moreover, our model incorporates the knowledge from the functional description of proteins extracted from literature or manual annotations.
Our experiments show that the S2F learns protein embeddings that achieve good performances on a variety of PPIs tasks, including cross-species PPI, antibody-antigen affinity prediction, antibody neutralization prediction for SARS-CoV-2, and mutation-driven binding affinity change prediction.
\end{abstract}


\section{Introduction}
Physical protein-protein interactions (PPIs) refer to the physical contacts between proteins due to interactions such as electrostatic forces, hydrogen bonding, and hydrophobic effects.
Many biological processes are driven by PPIs. For example, the severe acute respiratory syndrome coronavirus 2 (SARS-CoV-2) infects human cells starting from the PPI between the viral spike (S) protein and the receptor ACE2~\cite{hoffmann2020sars}.
Applications of PPIs range from vaccine design~\cite{polack2020safety,jackson2020mrna}, antibody therapeutics~\cite{shanmugaraj2020perspectives,lv2020structural}, to peptide drug discovery~\cite{bruzzoni2018interfering,marqus2017evaluation}.
Furthermore, in many applications, assessing the strength of the \textit{binding affinity} between proteins is important as well because the underlying biological mechanism involves the competitive binding between proteins.
Also, in the example of the SARS-Cov-2, the vaccinated person obtained the antibody binding to the N-terminal domain of the viral S protein to impede the original PPIs related to virus entry~\cite{polack2020safety}.
However, experimentally determining PPIs can be labor-intensive and time-consuming. Thus robust and accurate computational models are needed as an alternative approach.

\begin{figure}[ht!]
  \centering
  \includegraphics[width=0.8\columnwidth]{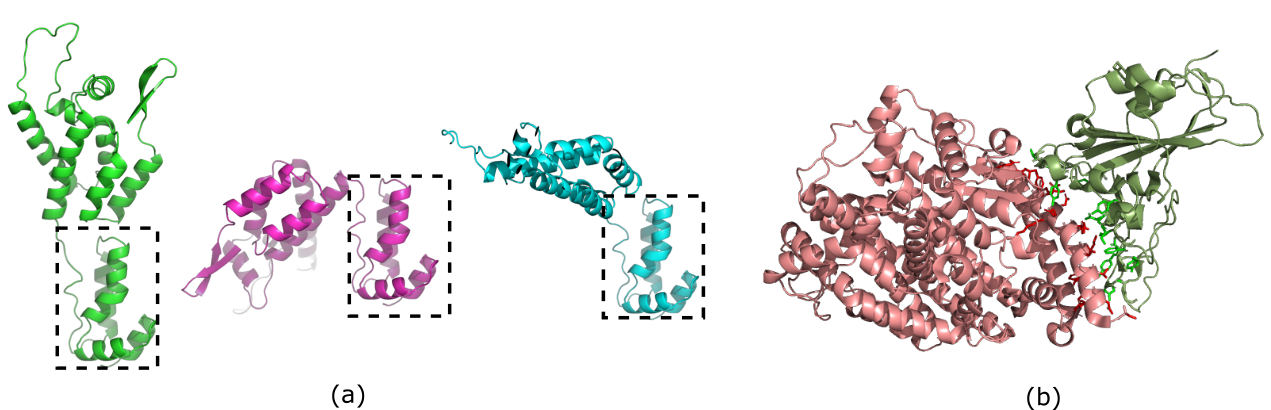}
  \caption{The missing but important information for PPIs prediction of existing protein encoding models. (\textbf{a}) An illustration of three different folding patterns of an intrinsically disordered protein (IDP) (HIV-1 capsid, PDB: 2M8N) which can interact with different proteins. For clearness, we align all the C-terminal domains in the boxes vertically. (\textbf{b}) A demonstration of PPI where the interaction involves the side chains. The side chains within the interaction interface are shown with the stick representation in different colors (PDB: 6M0J).}
  \label{fig:lm_cannot}
\end{figure}

Most computational methods to predict PPIs can be briefly divided into two phases: the protein feature encoding and the predictive machine learning models for particular PPI tasks~\cite{khatun2020evolution}. Appropriate protein encoding is crucial to predict PPIs accurately.
The straightforward protein encoding method includes the one-hot embedding, or the k-mer of amino acid residue ~\cite{li2017sprint}.
Beyond the straightforward methods, other methods incorporate physicochemical features like the AAindex~\cite{kawashima2000aaindex}, and the coevolutionary information of proteins in the form of a positional-specific scoring matrix (PSSM)~\cite{murakami2014homology,zhai2017highly}.
More recently, using the protein language model to learning a protein representation is shown able to capture rich biological information~\cite{rives2021biological}, and has been applied in several PPIs models~\cite{zhou2020mutation,filipavicius2020pre,sledzieski2021sequence}. Protein language models model the probability of sequences which is found to imply the evolutionary pressure and originally used for homology search in proteomics~\cite{altschul1998iterated,bateman2004pfam}.
Evolutionary information is helpful to predict local structures (secondary structure, contact map) and properties like stability, fluorescence, subcellular localization etc~\cite{rao2019evaluating,elnaggar2021prottrans}. However, a model trained with only the sequences is not enough to fully characterize the structural information, such as variants of the protein folding patterns as shown in Figure~\ref{fig:lm_cannot} (a).

Recently, more attention has been paid on structure-aware protein pre-training models.
For example, the protein family is used to incorporate some weak structural information~\cite{min2021pre}.
Contact maps are also used as one pre-training task to learn the structure feature~\cite{bepler2019learning,bepler2021learning}.
However, contact maps only consider amino acid-level rigid structure (or $C_{\alpha}$ only), thus don't have explicit information on the side chains and only have limited information about protein-protein interaction.
The detailed local structures, the side chains, play a more important role than the backbone structure in the PPI problem.
As shown in Figure~\ref{fig:lm_cannot} (b), due to the physical scales, proteins interaction occurs mostly on the side chains, while contact maps cannot explicitly indicate the conformation of side chains. 
In addition, for the proteins with no fixed 3D structures, like the intrinsically disordered proteins (IDPs)~\cite{wright2015intrinsically}, the protein function cannot be characterized only by one fixed 3D structure, and a high-level function description will be needed.


To alleviate the above limitations, we propose a multimodal protein model, which tries to extract a better protein representation by jointly learning the sequence-structure-function (\textbf{S2F}) features of proteins, inspired by the recent multimodal pre-training models in natural language processing and computer vision~\cite{su2019vl}.
Besides the protein sequence, we also include structure and function information to learn the protein representation. 
For the structure encoding, instead of using the contact map, we utilize the point clouds of heavy atoms (C, N, O) that consist of both backbone atoms and side-chain atoms.
Furthermore, the topology complex is adopted to model the flexibility of protein structure and extract the structure features, inspired by the works of Guo~\cite{xia2014persistent,wang2020topology}.
Comparing to the contact map, the topology feature can model structures more precisely, especially for the side chains (See the analysis in the Appendix~\ref{app:cmap_vs_topo}).
For the function modality, we use the textual description from related literature or manual annotations.
The multimodal protein data is extracted from the protein domains of the CATH database~\cite{sillitoe2021cath}, and the proteins from the UniProtKB/Swiss-Prot database~\cite{boutet2007uniprotkb}.
For downstream PPI tasks, due to the limitation of structure and function data, we can mask the structure and function tokens to get the protein embedding with only protein sequences. 

Our experimental results show that comparing to conventional protein language models and contact map-based structure-aware pre-training models, our multimodal S2F representation beats them in various PPI tasks, including cross-species PPI, antibody-antigen affinity prediction, antibody neutralization prediction for SARS-CoV-2, and mutation-driven affinity change prediction.

\section{Related Work}


Currently, the mainstream of the protein encoding methods for PPIs is to use a pre-training protein model to learn a protein representation. These works include MuPIPR~\cite{zhou2020mutation}, PaccMann-Proteomics~\cite{filipavicius2020pre}, D-SCRIPT~\cite{sledzieski2021sequence} etc.
Specifically, the embedding model of MuPIPR inspired by ELMo~\cite{peters2018deep} pre-trains a protein language model using bidirectional LSTM on the STRING database~\cite{szklarczyk2017string}. 
PaccMann-Proteomics extends TAPE~\cite{rao2019evaluating} to pre-train on pairs of interacting proteins in the STRING database.
D-SCRIPT utilizes Bepler \& Berger's model~\cite{bepler2019learning} which incorporates structure information using auxiliary tasks including structural similarity prediction and residue-residue contact prediction. Recently, Bepler \& Berger further develop their work and learn the structure knowledge and the protein language model in a multitask fashion~\cite{bepler2021learning}.
For convenience, we use their project repository name ProSE to refer to their pre-training model.
With the observations from natural language processing, it is anticipated that a larger pre-training model should also work better.
Thus, in this work, we also investigate a super large protein model, ProtTrans~\cite{elnaggar2021prottrans}, which implements Google's T5~\cite{raffel2020exploring} on protein data with 3 billion parameters comparing to TAPE~\cite{rao2019evaluating} with 34 million parameters.

The above-mentioned methods only need protein sequences to predict the PPIs. There are also methods that directly use protein structure as inputs for the PPI task, such as TopNetTree~\cite{wang2020topology}, GeoPPI~\cite{liu2021deep}, MutaBind2~\cite{zhang2020mutabind2}.
TopNetTree utilizes the topological representation of proteins and auxiliary features extracted from PDB files as the inputs of gradient-boosting trees for the mutation-driven affinity change prediction task~\cite{wang2020topology}.
GeoPPI implements the idea of self-supervised pre-training on protein structures via introducing a side-chain perturbation reconstruction task.
MutaBind2 shows that sophisticated feature engineering using the information in the PDB file can improve the performance on the PPI task.
The major disadvantage of the structure-based PPI methods is that the protein structure data is relatively scarce, and many PPI tasks only have protein sequences data available. By employing multimodal pre-training, our model is able to obtain a representation from the sequence but containing structure and function information.

\section{Method}

\subsection{Sequence-Structure-Function Transformer Model}
\label{sec:s2f}

On the whole, the proposed sequence-structure-function (\textbf{S2F}) transformer model is a single-stream multimodal model, in which three types of input tokens are fed into a single transformer and distinguished by input type embeddings.


\textbf{Sequence}.
The sequence in the S2F model refers to the amino acid sequence of a protein. The raw amino acid sequence is first tokenized by a protein tokenizer then converted to the amino acids embeddings.

\textbf{Function}.
The function modality in this work is the textual function description. Like the sequence modality, the function description is first tokenized by a function tokenizer then converted to the function tokens embeddings. Considering the large vocabulary of protein functions, we use BPE~\cite{sennrich2016neural} to reduce the embedding matrix size.

\begin{figure*}[ht]
  \centering
  \includegraphics[width=\columnwidth]{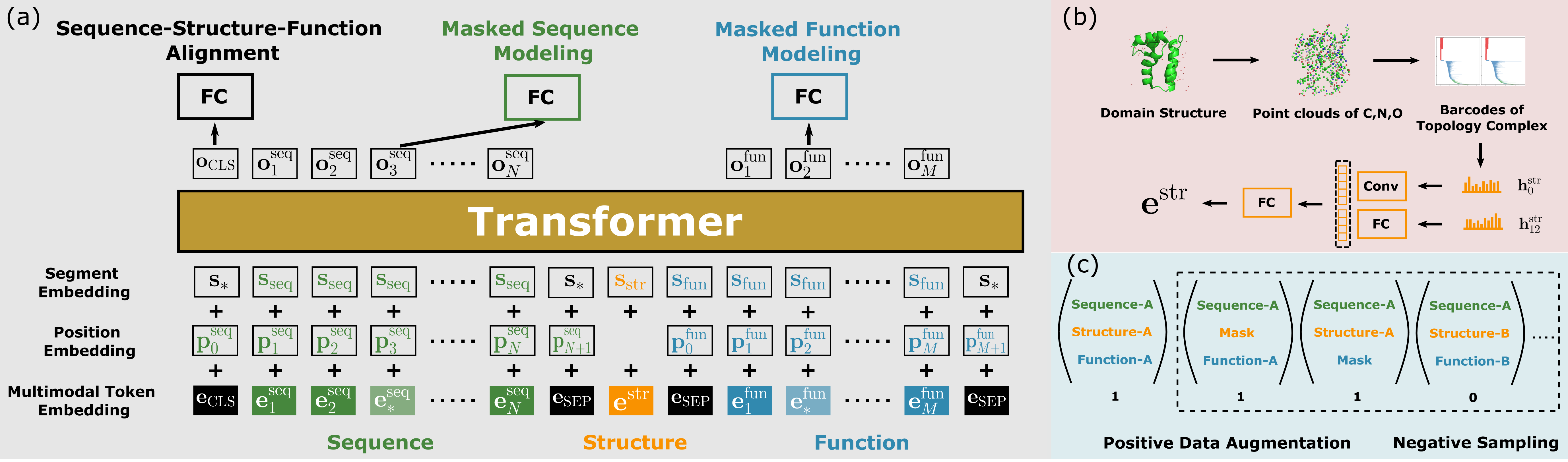}
  \caption{Overview of the Sequence-Structure-Function (S2F) protein pre-training model. Green, orange, and blue elements represent the sequence, structure, and function modality respectively. (\textbf{a}) The architecture of the single-stream multimodal transformer model and three pre-training tasks: sequence-structure-function alignment, masked sequence modeling and masked function modeling. Note that $\mathbf{e}_{*}^{\mathrm{seq}}$ and $\mathbf{e}_{*}^{\mathrm{fun}}$ mean the masked tokens. (\textbf{b}) The pipeline of the topology encoder that extracts the topological structure token $\mathbf{e}^{\mathrm{str}}$ from the domain structure. (\textbf{c}) An illustration of the positive data augmentation and negative sampling processes for pre-training tasks.}
  \label{fig:s2f_overview}
\end{figure*}

\textbf{Structure}.
Our S2F model utilizes a topology encoder adopting TopNetTree~\cite{wang2020topology} to get the structure embedding.
Specifically, the vectorized Rips complex barcodes and Alpha complex barcodes are feed into a convolutional network and fully connected network respectively, then concatenate them and apply a joint fully connected network to learn the ensemble structure embedding (Figure~\ref{fig:s2f_overview}(b)).
If we denote the point cloud generated by A-atom and B-atom as $\mathbf{P}_{A,B}$, the Rips barcodes calculation and vectorization by $\Rips(\mathbf{P}_{A,B})$, and the Alpha barcodes calculation and vectorization by $\Alpha(\mathbf{P}_{A,B})$, the progress of topology structure encoding can be represented by the following equations:

\begin{align}\label{eq:topo_encoder}
  \begin{split}
    \mathbf{h}_0^{\mathrm{str}} &= \Concat\limits_{A, B \in \{C, N, O\}} {\Rips(\mathbf{P}_{A,B})}, \\
    \mathbf{h}_{12}^{\mathrm{str}} &= \Concat\limits_{A, B \in \{C, N, O\}} {\Alpha(\mathbf{P}_{A,B})}, \\
    \mathbf{e}^{\mathrm{str}} &= \Concat \left(\FC_2(\Conv(\mathbf{h}_0^{\mathrm{str}}), \FC_1(\mathbf{h}_{12}^{\mathrm{str}}))\right),
  \end{split}
\end{align}
where $\FC(\cdot)$ is a fully connected network, $\Conv(\cdot)$ is a convolutional network, and $\mathbf{e}^{\mathrm{str}}$ is the topological structure token.

Like BERT~\cite{kenton2019bert}, for each input token, the positional embedding is added such that the transformer model can have the sequential information.
In our model, we apply two different positional embeddings for the sequence and function tokens and attach a segment embedding to indicate different types of tokens.
This final inputs for the transformer can be obtained from Equation~\eqref{eq:seqTokens} to ~\eqref{eq:funTokens},

\begin{align}
  \begin{split}\label{eq:seqTokens}
    \mathbf{x}_{\mathrm{CLS}} &= \mathbf{e}_{\mathrm{CLS}} + \mathbf{p}_0^{\mathrm{seq}} + \mathbf{s}_{*} \\
    \mathbf{x}_i^{\mathrm{seq}} &= \mathbf{e}_i^{\mathrm{seq}} + \mathbf{p}_i^{\mathrm{seq}} + \mathbf{s}_{\mathrm{seq}}, \quad i=1,\ldots,N\\
    \mathbf{x}_{\mathrm{SEP},1} &= \mathbf{e}_{\mathrm{SEP}} + \mathbf{p}_{N+1}^{\mathrm{seq}} + \mathbf{s}_{*} \\
  \end{split} \\[5pt]
  \begin{split}\label{eq:strTokens}
    \quad\; \mathbf{x}^{\mathrm{str}} &= \mathbf{e}^{\mathrm{str}} + \mathbf{s}_{\mathrm{str}} \\
  \end{split}\\[5pt]
  \begin{split}\label{eq:funTokens}
    \mathbf{x}_{\mathrm{SEP},2} &= \mathbf{e}_{\mathrm{SEP}} + \mathbf{p}_0^{\mathrm{fun}} + \mathbf{s}_{*} \\
    \mathbf{x}_j^{\mathrm{fun}} &= \mathbf{e}_j^{\mathrm{fun}} + \mathbf{p}_j^{\mathrm{fun}} + \mathbf{s}_{\mathrm{fun}}, \quad j=1,\ldots,M\\
    \mathbf{x}_{\mathrm{SEP},3} &= \mathbf{e}_{\mathrm{SEP}} + \mathbf{p}_{M+1}^{\mathrm{fun}} + \mathbf{s}_{*}, \\
  \end{split}
\end{align}

where $\mathbf{e}_i^{\mathrm{seq}}$ is the $i$-th amino acid token, $\mathbf{e}_j^{\mathrm{fun}}$ is the $j$-th subword token of function, $\mathbf{e}_{\mathrm{CLS}}$ is the classifier token, $\mathbf{e}_{\mathrm{SEP}}$ is the separator token, $\mathbf{p}_i^{\mathrm{seq}}$ and $\mathbf{p}_j^{\mathrm{fun}}$ are positional embeddings of sequence and function tokens respectively, $N$ is the length of sequence, $M$ is the length of the function description, $\mathbf{s}_{*}$, $\mathbf{s}_{\mathrm{seq}}$, $\mathbf{s}_{\mathrm{str}}$ and $\mathbf{s}_{\mathrm{fun}}$ are the segment embeddings to indicate four types of tokens.
These input tokens are converted to feature vectors via a transformer network.
In the pre-training phase, the outputs of the transformer are the feature vectors derived from the classifier token, the sequence tokens, and the function tokens, which can be represented by
\begin{align}\label{eq:transformer}
  \begin{split}
  & \quad \; \mathbf{o}_{\mathrm{CLS}}, \mathbf{o}_1^{\mathrm{seq}}, \ldots, \mathbf{o}_N^{\mathrm{seq}}, \mathbf{o}_1^{\mathrm{fun}}, \ldots, \mathbf{o}_M^{\mathrm{fun}} \\ &= \TF \left(
  \mathbf{x}_{\mathrm{CLS}}, \mathbf{x}_1^{\mathrm{seq}}, \ldots, \mathbf{x}_N^{\mathrm{seq}}, \mathbf{x}_{\mathrm{SEP},1}, \mathbf{x}^{\mathrm{str}}, \mathbf{x}_{\mathrm{SEP},2}, \mathbf{x}_1^{\mathrm{fun}}, \ldots, \mathbf{x}_M^{\mathrm{fun}}, \mathbf{x}_{\mathrm{SEP},3}\right).
  \end{split}
\end{align}




\subsection{Pre-training tasks}
We use three tasks when pre-training the S2F model (Figure~\ref{fig:s2f_overview}). The first is the sequence-structure-function alignment, which is inspired by the previous work on the visual-text pre-training model~\cite{su2019vl}.
In this task, the model learns to figure out whether the input tokens of different modalities are from the same example (Figure~\ref{fig:s2f_overview}(c)).
In the context of our S2F model, the model makes decisions using its \textbf{CLS} header to indicate whether the topological structure encoding corresponds to a specific amino acid sequence or whether the protein function matches the amino acid sequence.
Aligning a sequence with its structure topology helps the model to encode meaningful structural information as there are already some works on deciphering protein sequences to low-level structures~\cite{rao2019evaluating,elnaggar2021prottrans}.
Although the gap between a protein sequence and its function might be huge at first glance, we argue that even simply learning the co-occurrence probability of the protein sequence and its scientific function tokens is meaningful. The function descriptions contain important information about PPIs, such as where the protein acts and what other protein it cooperates with.
Moreover, thanks to the growing biological research, we can collect more such function data and learn the newest biological knowledge reforming the current model radically.

The other two tasks are variants of the conventional masked language modeling~\cite{kenton2019bert}, a self-supervised method on the sequential data.
We randomly mask out individual residue tokens within the protein sequences or word tokens within the function texts and use a BERT-style objective to predict the masked tokens. The BERT objective is trained jointly with the above alignment task.
We apply fully connected neural networks on the outputs of the transformer network to get the predictions of the pre-training tasks. The total training loss of the above three tasks can be written as Equation~\eqref{eq:loss}

\begin{equation}\label{eq:loss}
  \begin{gathered}
    \mathcal{L} = \mathcal{L}_{\text{ce}}(y_{\text{cls}}, \FC_3(\mathbf{o}_{\mathrm{CLS}})) + \sum_{i \in \text{Masked}}\mathcal{L}_{\text{ce}}(y_i^{\mathrm{seq}}, \FC_4(\mathbf{o}_i^{\mathrm{seq}})) + \sum_{j \in \text{Masked}} \mathcal{L}_{\text{ce}}(y_j^{\mathrm{fun}}, \FC_5(\mathbf{o}_j^{\mathrm{fun}})),
  \end{gathered}
\end{equation}

where $\mathcal{L}_{\text{ce}}(\cdot)$ is the cross-entropy loss, $y_{\text{cls}}$ is the 0-1 label indicating whether cross modalities are matched, $y_i^{\mathrm{seq}}$ is the class id (i.e. types of amino acids) of the masked $i$-th sequence token, and $y_j^{\mathrm{fun}}$ is the class id for the masked $j$-th function token.

\subsection{Downstream PPI Tasks}
\label{sec:ppi_tasks}
\begin{figure*}[ht]
  \centering
  \includegraphics[width=\columnwidth]{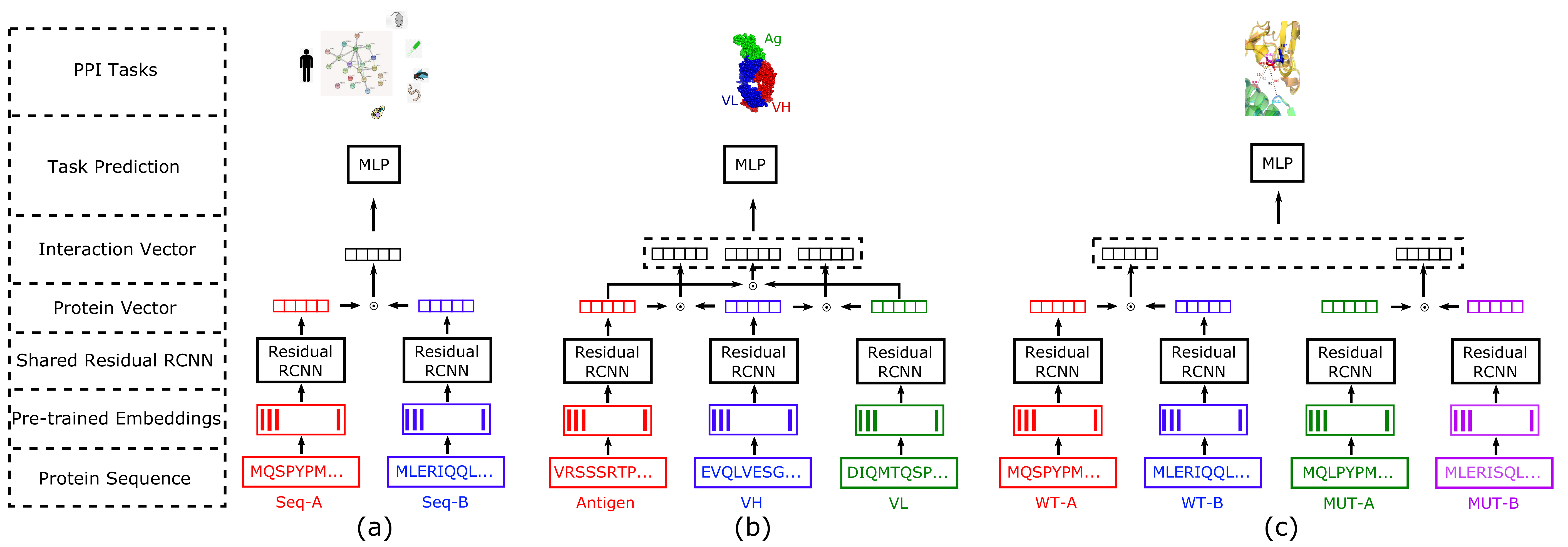}
  \caption{Variants of PIPR-like networks~\cite{chen2019multifaceted} for downstream PPI tasks. (\textbf{a}) The network for the cross-species PPI task. (\textbf{b}) The network for antibody-antigen interaction tasks. (\textbf{c}) The network for mutation-driven affinity change prediction task. Note that the weights of the Residual RCNN module are shared in each specific task. $\odot$ means element-wise multiplication.}
  \label{fig:downstream_tasks}
\end{figure*}

Since we aim at developing a sequence-based PPI model and there are no structure or function data in many PPI problems, we use the \textbf{MASK} tokens as the structure and the function tokens when applying the pre-training model, which means we replace $\mathbf{e}^{\mathrm{str}}$ and $\mathbf{e}_j^{\mathrm{fun}}$ with the \textbf{MASK} tokens. 
After pre-training the S2F model, we take the feature vectors from the sequence ($\mathbf{o}_1^{\mathrm{seq}},\ldots,\mathbf{o}_N^{\mathrm{seq}}$ in Equation~\eqref{eq:transformer}) as the embedding of the protein, which is the \textbf{pre-trained embeddings} for the downstream PPI prediction models as shown in Figure~\ref{fig:downstream_tasks}.


On top of the protein representation learned by S2F, we utilize different neural networks based on the Residual RCNN from PIPR~\cite{chen2019multifaceted} to make the PPI prediction, according to different kinds of PPI tasks: cross-species PPI, antibody-antigen interaction, and mutation-driven affinity change prediction (Figure~\ref{fig:downstream_tasks}). 

\textbf{Cross-species PPI}. In this task, we train a binary classifier on the human PPI data, then directly evaluate its generalization capability on new species, including worm, fly, yeast, E.coli, and mouse.
The network architecture for this task is shown as Figure~\ref{fig:downstream_tasks}(a).
As a canonical PPI task, it accepts two protein sequences as input, uses element-wise multiplication to mimic the interaction process, and applies the binary cross-entropy loss to train the model.

\textbf{Antibody-Antigen interaction}. In this work, we only study the cases that the antigen is a protein or a peptide. In such cases, the protein interactions are usually associated with three or more protein sequences. We consider two similar tasks here: the antibody-antigen affinity prediction and the antibody neutralization prediction.
For the antibody-antigen affinity prediction task, we need to take the Fab domain (VH and VL chains) and the antigen into considerations.
While in the antibody neutralization prediction task, it has an additional receptor protein that is the target of the antigen. In this work, we study the SARS-CoV-2 antibody neutralization task, which has a common receptor protein, such that it can share the same network architecture as the antibody-antigen affinity prediction task.
The overall network architecture is illustrated in Figure~\ref{fig:downstream_tasks}(b).
We use the mean square error loss for the affinity prediction task and the binary cross-entropy loss for the antibody neutralization prediction task.

\textbf{Mutation-driven affinity change prediction}. Mutations can dramatically change the properties of a protein by changing the binding affinity of PPIs. In this task, the network accepts the sequence pairs before and after the mutations and tries to predict the affinity change represented by free energy change, $\Delta \Delta G$. We use a network architecture similar to MuPIPR~\cite{zhou2020mutation}, as shown in Figure~\ref{fig:downstream_tasks}(c).
It is a regression task, so the mean square error loss is used to train the network.

\section{Results}

\subsection{Multi-modality Protein Data for Pre-Training}
In order to perform the sequence-structure-function pre-training, we compile a multi-modality protein dataset from two databases, the CATH Protein Structure Classification database~\cite{sillitoe2021cath}\footnote{In this work, we downloaded the RCSB PDB database at 2021.3.15, \href{http://www.rcsb.org/}{http://www.rcsb.org/}} and the reviewed section of the UniProt Knowledgebase, UniProtKB/Swiss-Prot~\cite{boutet2007uniprotkb}\footnote{We collected UniProtKB/Swiss-Prot 2021-03 Release, \href{https://www.uniprot.org/downloads}{https://www.uniprot.org/downloads}}.
Specifically, for the structure data, we use the structures of protein domains because they are relatively stable and independent components comparing to the whole protein.
We use PyMOL~\cite{PyMOL} to purify and extract domain PDB files from RCSB Protein Data Bank~\cite{burley2021rcsb} based on the CATH annotations. The CATH annotations are also used to train a topology feature classifier, which is used to initialize the weights for the neural networks in Equation~\eqref{eq:topo_encoder}. 
For the function data, we collect the function entry from the Swiss-Prot database and the function of the CATH codes of the protein domain.
Statistically, we collected 258,928 domains data in sequence-structure-function format, and 564,638 proteins data in sequence-function format.
A more detailed description of the data cleaning and processing can be found in Appendix~\ref{app:pretrain_data}.

\subsection{Cross-species PPI}
We first use a canonical PPI task to validate that the S2F learned embeddings are helpful to the PPI prediction.
In the cross-species PPI task, we train a PPI classifier using a large PPI dataset on human and evaluate its performance on small PPI datasets from 5 different species: mouse, fly, worm, yeast, and E.coli.
Specifically, the human PPI dataset has around 420K protein pairs. While for the five testing species, the E.coli dataset has only 22K pairs, and all other datasets have 55K protein pairs.
All of these PPI datasets are obtained from Sledzieski et al.'s recent work and have a positive-to-negative ratio 1:10~\cite{sledzieski2021sequence}.

Table~\ref{tab:cs_ppi} compares the performances of PIPR with S2F embeddings and the original PIPR, which uses one-hot encoding and the physicochemical features to embed the proteins~\cite{chen2019multifaceted}. 
The PIPR network with the S2F multimodal model embeddings significantly outperforms the original PIPR on five testing species.
The experimental results of this cross-species PPI task show that the S2F learned embeddings can greatly enhance the representation and have a better generalization ability.
Here, it is worth mentioning that for all of our downstream PPI tasks, no structure or function information is needed as inputs because we can use \textbf{MASK} tokens as the placeholders (See Section~\ref{sec:ppi_tasks} for details).


\begin{table}[ht]
\centering
\caption{Results on Cross-Species PPI}
\label{tab:cs_ppi}
\begin{tabular}{|c|c|c|c|c|c|}
\hline
Species                                                                      & Methods  & AUPR           & Precision      & Recall         & AUROC          \\ \hline
\multirow{2}{*}{Mouse}                                                       & PIPR     & 0.526          & 0.734          & 0.331          & 0.839          \\ \cline{2-6} 
                                                                             & PIPR+S2F & \textbf{0.644} & \textbf{0.781} & \textbf{0.447} & \textbf{0.889} \\ \hline
\multirow{2}{*}{Fly}                                                         & PIPR     & 0.278          & 0.521          & 0.121          & 0.728          \\ \cline{2-6} 
                                                                             & PIPR+S2F & \textbf{0.450} & \textbf{0.652} & \textbf{0.286} & \textbf{0.795} \\ \hline
\multirow{2}{*}{Worm}                                                        & PIPR     & 0.346          & 0.673          & 0.142          & 0.757          \\ \cline{2-6} 
                                                                             & PIPR+S2F & \textbf{0.484} & \textbf{0.724} & \textbf{0.270} & \textbf{0.825} \\ \hline
\multirow{2}{*}{Yeast}                                                       & PIPR     & 0.230          & 0.398          & 0.085          & 0.718          \\ \cline{2-6} 
                                                                             & PIPR+S2F & \textbf{0.356} & \textbf{0.604} & \textbf{0.180} & \textbf{0.767} \\ \hline
\multirow{2}{*}{E. coli}                                                     & PIPR     & 0.308          & \textbf{0.629} & 0.131          & 0.675          \\ \cline{2-6} 
                                                                             & PIPR+S2F & \textbf{0.371} & 0.467          & \textbf{0.309} & \textbf{0.748} \\ \hline
\multirow{2}{*}{\begin{tabular}[c]{@{}c@{}}Human\\ (5-fold CV)\end{tabular}} & PIPR     & \textbf{0.835} & \textbf{0.838} & \textbf{0.701} & \textbf{0.960} \\ \cline{2-6} 
                                                                             & PIPR+S2F & 0.822          & 0.836          & 0.672          & 0.956          \\ \hline
\end{tabular}
\end{table}

\subsection{Antibody-antigen Interaction}
We then compare the proposed S2F embeddings with various existing protein pre-training models, and the overall information is listed in Table~\ref{tab:protein_models}. 
Regarding the training data size, the smallest one is the ELMo model~\cite{zhou2020mutation}, which only trains on protein sequences of \textit{Homo sapiens}, \textit{Bos taurus}, \textit{Mus musculus}, and \textit{Escherichia coli} in the STRING database~\cite{szklarczyk2017string}; the ProSE model~\cite{bepler2021learning} uses the largest training data size.
Notably, ProSE incorporates structure information in form of contact maps and the structure classification from the SCOP database~\cite{murzin1995scop}.
With respect to the network architectures, TAPE~\cite{rao2019evaluating}, ProtTrans~\cite{elnaggar2021prottrans}, and our S2F are based on transformers; ELMo~\cite{zhou2020mutation} and ProSE~\cite{bepler2021learning} are based on bidirectional LSTM.


\begin{table}[!ht]
\centering
\caption{Details of Different Pre-trained Protein Models}
\label{tab:protein_models}
\begin{tabular}{|c|c|c|c|c|}
\hline
Protein Model & Pre-training Dataset     & Data Size & Parameters & Embedding Size \\ \hline
ELMo~\cite{zhou2020mutation}                      & Subset of STRING         & 70K       & 200K       & 128            \\ \hline
TAPE~\cite{rao2019evaluating}                      & Pfam                     & 31M       & 34M        & 512            \\ \hline
ProtTrans~\cite{elnaggar2021prottrans}                 & Uniref50                 & 45M       & 3B         & 1024           \\ \hline
ProSE~\cite{bepler2021learning}                     & SCOP+UniRef90            & 28K+76M   & 97M        & 6165           \\ \hline
S2F (Ours)                 & CATH+UniProtKB/SwissProt & 260K+560K & 34M        & 512            \\ \hline
\end{tabular}
\end{table}

In this subsection, we consider the problem of antibody-antigen interaction, which is an important kind of PPIs in life processes. To evaluate our proposed pre-trained embeddings, we use two antibody-antigen tasks:
\begin{itemize}
    \item \textbf{SAbDab}, an antibody-antigen affinity prediction task. The dataset is from the SAbDab database~\cite{dunbar2014sabdab} with 494 examples. Each example has an antigen sequence, a variable heavy chain, a variable light chain, and the affinity label.
    \item \textbf{SARS-CoV-2 Antibody Neutralization}, an antibody neutralization prediction task, to predict whether the given antibody can neutralize the effect of S protein and hinder the infection of SARS-CoV-2. The dataset is from the CoV-AbDab database~\cite{raybould2021cov} with 747 positive examples and 330 negative examples. Note that the negative examples are selected from the antibodies that can bind to the receptor binding domain (RBD) of the S protein in SARS-CoV-2 but cannot neutralize the virus, which makes the task even harder.
\end{itemize}

For the antibody-antigen affinity prediction task, SAbDab, we use the root-mean-square error (RMSE) and Pearson's correlation (Rp) to evaluate the performance. For the classification task, SARS-CoV-2 Antibody Neutralization, we use AUROC, AUPR, F1 score, precision, and recall to evaluate the model. 10-fold cross-validation is performed, and the average across all folds is reported in Table~\ref{tab:antibody_antigen}. It is easy to see that the proposed S2F model beats the other pre-trained embeddings in both regression and classification tasks.

\begin{table}[ht]
\centering
\caption{Results on Antibody-Antigen Tasks}
\label{tab:antibody_antigen}
\begin{tabular}{|c|c|c|c|c|c|c|c|}
\hline
\multirow{2}{*}{Methods} & \multicolumn{2}{c|}{SAbDab}     & \multicolumn{5}{c|}{SARS-CoV-2 Antibody Neutralization}                             \\ \cline{2-8} 
                         & RMSE           & Rp             & Precision      & Recall         & F1             & AUPR           & AUROC          \\ \hline
PIPR+ELMo                & 0.304          & 0.317          & \textbf{0.745} & 0.657          & 0.698          & 0.792          & 0.603          \\ \hline
PIPR+TAPE                & 0.277          & 0.517          & 0.735          & 0.788          & 0.761          & 0.801          & 0.637          \\ \hline
PIPR+ProtTrans           & 0.274          & 0.521          & 0.728          & 0.807          & 0.766          & 0.794          & 0.623          \\ \hline
PIPR+ProSE               & 0.273          & 0.518          & 0.721          & 0.704          & 0.712          & 0.794          & 0.612          \\ \hline
PIPR+S2F (Ours)           & \textbf{0.272} & \textbf{0.526} & 0.742          & \textbf{0.861} & \textbf{0.797} & \textbf{0.836} & \textbf{0.693} \\ \hline
\end{tabular}
\end{table}

\subsection{Mutation-driven Affinity Change Prediction}

Besides antibody-antigen recognition, other important roles of PPIs include virus entry and tumor cell immune escape. Considering these two functions become unpredictable and easy-perturbed by mutations, in this work, we also test the proposed S2F model on the mutation-driven affinity change prediction task.
We perform the experiments on two datasets for this task:

\begin{itemize}
  \item \textbf{S645}~\cite{pires2016mcsm}, an affinity change dataset for single-site mutations. The dataset is compiled from the AB-Bind database~\cite{sirin2016ab}, with a total of 645 single-point mutations on 29 different Ab–antigen complexes.
  \item \textbf{M1535}, an affinity change dataset for multiple sites mutations. The dataset is filtered from SKEMPI 2.0 database~\cite{jankauskaite2019skempi}. It is a subset of the M1707 dataset with only two sequences in the PDB files~\cite{zhang2020mutabind2}.
\end{itemize}

The samples in the above two datasets are mutations, and the sequences may be very similar if we randomly split them. To minimize the effect of over-fitting and test the ability of generalization, we divide the collected datasets into groups that are different in both structure and homology, with the homology code from the ECOD database~\cite{cheng2014ecod} and the structure id from the RCSB PDB database~\cite{burley2021rcsb}. We then merge these initial clusters to organize a 5-fold split and make them as even as possible. We report the performance of different pre-training models with cross-validation on these 5-fold splits.
We call such an evaluation method as \textbf{Homology\&Structure-split Cross Validation} (HSCV).
As shown in Table~\ref{tab:ddG}, the HSCV evaluation makes the task more challenging, and the proposed S2F embedding achieves the best performance in terms of both RMSE and Pearson's correlation.

\begin{table}[ht]
\centering
\caption{Results on Mutation-driven Affinity Change Prediction Tasks using Homology\&Structure-Split Cross Validation}
\label{tab:ddG}
\begin{tabular}{|c|c|c|c|c|}
\hline
\multirow{2}{*}{Methods} & \multicolumn{2}{c|}{S645-HSCV}  & \multicolumn{2}{c|}{M1535-HSCV} \\ \cline{2-5} 
                         & RMSE           & Rp             & RMSE           & Rp             \\ \hline
PIPR+ELMo                & 1.913          & 0.289          & 3.229          & 0.220          \\ \hline
PIPR+TAPE                & 1.895          & 0.273          & 3.260          & 0.155          \\ \hline
PIPR+ProtTrans           & 1.888          & 0.284          & 3.240          & 0.196          \\ \hline
PIPR+ProSE               & 1.912          & 0.257          & 3.279          & 0.142          \\ \hline
PIPR+S2F (Ours)           & \textbf{1.866} & \textbf{0.318} & \textbf{3.199} & \textbf{0.264} \\ \hline
\end{tabular}
\end{table}

\section{Discussion}

In this work, we introduce a multimodal protein pre-training model that incorporates sequence, structure, and function data of the protein to jointly train a protein representation for the PPI problem.
Our idea is inspired by a more holistic view of proteins that interweaves sequence, structure, and function.
By masking the structure and function related input tokens, we can adopt the proposed pre-training model for sequence-based PPI tasks.
We assess our pre-trained protein embeddings on various PPI tasks. The first experiment on cross-species PPI shows that our pre-training protein embeddings outperform traditional one-hot and physicochemical feature-based representations.
We also compare with several other protein pre-training embeddings. With the same downstream network, the experimental results show that our multimodal pre-training embeddings can boost the performance on different PPI tasks.

From the aspect of applications, we can easily adapt the protein embeddings learned by S2F to other protein problems, such as protein annotation and protein stability prediction. For future works, it is possible to use the high-quality computed protein structures from AlphaFold2 to increase the multimodality data size for pre-training our model~\cite{jumper2021highly}. Besides, one may use a pre-trained language model to encode the textual function data better for the function modality. We hope to learn richer knowledge and build better representation for proteins via larger and better datasets, then applying the representation to PPI-related drug discovery.

\bibliographystyle{plain}
\bibliography{./ref}  

\newpage
\begin{appendices}

\section{Pre-training Data Preparation}
\label{app:pretrain_data}

\textbf{Domain data processing}.
Extracting structures and functions of domains in the PDB file involves several steps.
First, we remove solvent molecules and ligands in the PDB file. Because these molecules also have heavy atoms C, N, O, and unwanted heavy atoms will be included in the point clouds when we compute the topology complex features.
We then split protein chains and remove the repeated structures.
Finally, we apply sequence alignment over each protein sequence and domains sequences in the CATH database~\cite{sillitoe2021cath} to get the corresponding CATH code and function description.

\textbf{Topology barcodes extraction and vectorization}.
Once the domains are extracted, we compute the topology complex features from the point clouds of the heavy atoms (C, N, O).
Inspired by TopNetTree~\cite{wang2020topology}, we use both Alpha complex and Rips complex.
Specifically, for each heavy atom pair A-B (one of C-C, N-N, O-O, C-N, C-O, N-O), we calculate the barcodes of the Rips and Alpha complex using the distance matrix of the point cloud represented by all A-atoms and B-atoms.
Considering the physical size of protein domains, we select the cutoff 0.01 angstrom and 20 angstroms, meaning that the barcodes with length (death minus birth) smaller than 0.01 angstrom or with death value larger than 20 angstroms are ignored.
Then we make bins to count the number of barcodes with interval 0.25 angstrom.
These bins normalized by the mean and the standard deviation over 258,928 examples create a vector as a low-dimensional structure topology representation.

\textbf{Topology feature classifier}.
Since the CATH codes describe a protein in four levels (class, architecture, topology, homologous superfamily), we treat them as classification labels of proteins.
Using these labels and the topology encoder network (Figure~\ref{fig:s2f_overview} (b)), we can train a classifier that can act as good initialization weights for the proposed pre-training model. 
Statistically, there are 3,763 distinct CATH codes within our domains data.

\section{Contact Map and Topology Feature for IDPs}
\label{app:cmap_vs_topo}

\begin{figure}[ht]
  \centering
  \includegraphics[width=0.6\columnwidth]{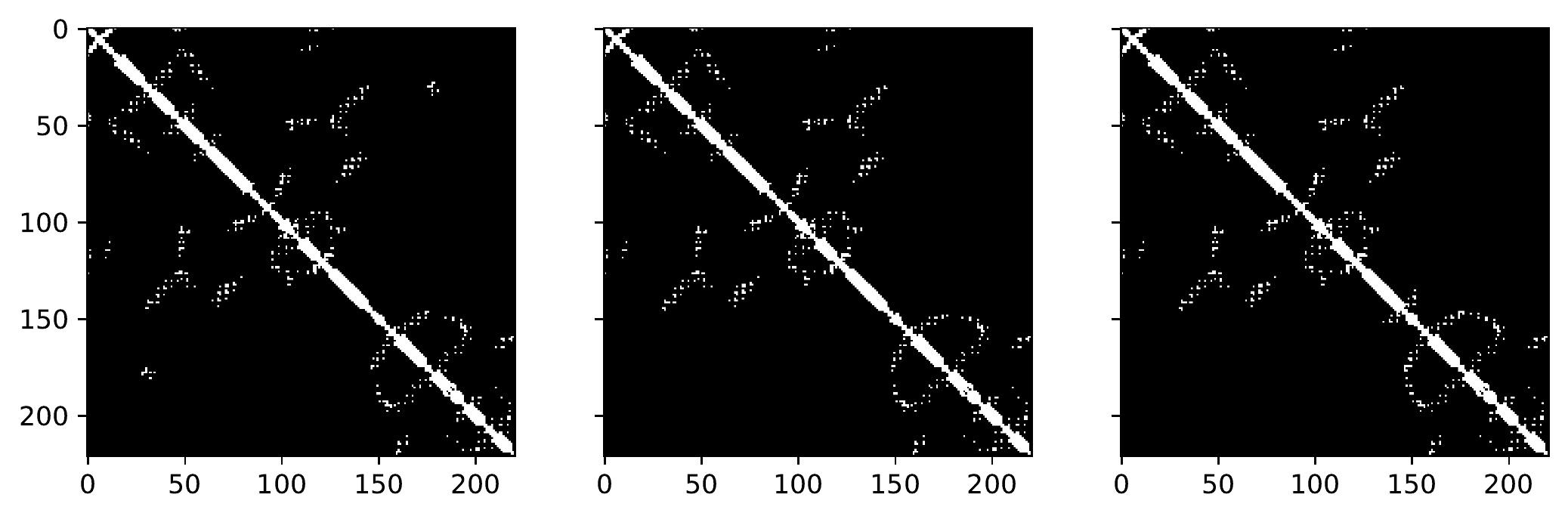}
  \caption{Contact maps of HIV-1 capsid protein with different folding patterns. Note that the threshold of the contact map is 8 angstroms.}
  \label{fig:cmap}
\end{figure}

Intrinsically disordered proteins (IDPs) allow the same protein to undertake different interactions under different folding patterns. Therefore, IDPs act as important components of cellular signaling machinery and usually have a large degree in the PPI networks~\cite{wright2015intrinsically}.
To better predict the PPIs related to IDPs, there is an unmet need to model their variants of structures precisely.
Most existing works model the protein structure at the level of contact maps. However, for IDPs, different folding patterns cannot be precisely presented by contact map because of its characteristic of locality.
For example, the contact maps in Figure~\ref{fig:cmap} are corresponding to the three different folding patterns of HIV-1 capsid protein in Figure~\ref{fig:lm_cannot} (a).
It is hard to characterize the significantly different folding patterns from the nuances between the contact maps.

\begin{figure}[ht]
  \centering
  \includegraphics[width=0.8\columnwidth]{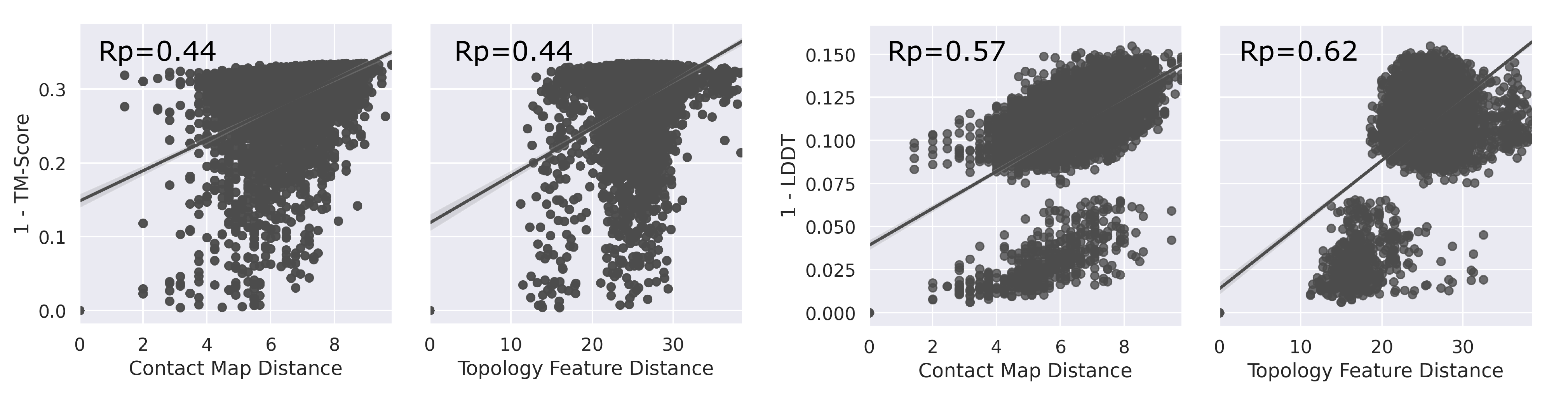}
  \caption{Correlations of $1-\text{TMScore}$ and $1-\text{LDDT}$ with the contact map distance and the topology feature distance.}
  \label{fig:cmap_vs_topo}
\end{figure}

Furthermore, to assess the different capabilities of contact map and topology feature at modeling the side chains, we examine the correlations between the structural differences and the distances given by different structure encoding vectors.
Specifically, we collect 100 different structures of the HIV-1 capsid protein from the PDB file 2m8n. Then we calculate the pairwise structure differences using TM-Score~\cite{zhang2004scoring} and LDDT~\cite{mariani2013lddt}.
For both TM-Score and LDDT, the higher score (more closer to 1) means higher structure similarity.
Since TM-Score only considers the $C_{\alpha}$ and LDDT scores are computed on all atoms, we can use $1 - \text{TMScore}$ as the structure difference without the side chains, and use $1 - \text{LDDT}$ as the structure difference with the side chains.
Next, contact maps are flattened as encoding vectors and we calculate the pairwise Euclidean distances as the \textbf{contact map distance}. Similarly, we can get the \textbf{topology feature distance} from the normalized topology feature vectors.
Finally, we compute two pairs of correlations shown in Figure~\ref{fig:cmap_vs_topo}. 
As we can see, without the side chains information, both the contact map distance and the topology feature distance have the same Pearson correlation with the structural difference.
While when considering the side chains, the topology feature distance has a higher correlation with the structure differences than the contact map distance.
This result indicates that the topology feature can model the variants of structures with side chains more precisely.

\end{appendices}

\end{document}